\documentclass[aps,prl,twocolumn,groupedaddress,showpacs,showkeys,floatfix,superscriptaddress]{revtex4-1}

\usepackage{graphicx}
\usepackage{epsfig}

\usepackage{amssymb}
\usepackage{amsmath}
\usepackage{color}

\begin{document}

\title{Probing the electromagnetic local density of states \\with a strongly mixed electric and magnetic dipole emitter}

\author{Sinan Karaveli}
\altaffiliation{Present address: Research Laboratory of Electronics, Massachusetts Institute of Technology, Cambridge, MA 02139, USA}
\address{School of Engineering, Brown University, Providence, RI 02912, USA}
\author{Dongfang Li}
\address{School of Engineering, Brown University, Providence, RI 02912, USA}
\address{Department of Physics, Brown University, Providence, RI 02912, USA}
\author{Rashid Zia}
\email{Rashid\_Zia@brown.edu}
\address{School of Engineering, Brown University, Providence, RI 02912, USA}
\address{Department of Physics, Brown University, Providence, RI 02912, USA}

\date{\today}

\begin{abstract}
We identify a solid-state quantum emitter whose room-temperature radiative decay is mediated by a nearly equal mixture of isotropic electric dipole (ED) and magnetic dipole (MD) transitions. Using energy-momentum spectroscopy, we experimentally show that the near-infrared $^3$T$_2{\rightarrow}^3$A$_2$  emission from divalent-nickel-doped magnesium oxide (Ni$^{2+}$:MgO) is composed of $\sim$50\% MD and $\sim$50\% ED transitions. We then demonstrate that the spontaneous emission rate of these ions near planar interfaces is determined by the combined electric and magnetic local density of optical states (LDOS). This electromagnetic LDOS probes the total mode density, and thus similar to thermal emission, these unique electronic emitters effectively excite all polarizations and orientations of the electromagnetic field. 

\end{abstract}
\pacs{32.50.+d, 32.70.-n, 42.50.Ct, 78.66.-w}

\maketitle
The advent of metamaterials has helped to highlight the diversity of electromagnetic resonances in optical nanostructures. For example, recent studies have shown how the interplay of electric and magnetic resonances can enable exciting new effects, such as zero optical backscattering~\cite{Geffrin, FuNatComm, Person}, enhanced chiral spectroscopy~\cite{Etxarri2}, and robust optical activity~\cite{Sersic1}. The interaction of electric and magnetic resonances has also been used to reproduce phenomena from atomic physics, e.g. Fano resonances ~\cite{Fan,EvlyukhinPRB,Sheikholeslami,Luk'yanchuk}. In this regard, there are strong similarities between the lifetime of microscopic quantum emitters and the linewidth of macroscopic scatterers~\cite{Buchler}, but there are also intriguing differences. Recent calculations have revealed that the scattering linewidth of a split-ring resonator should follow a newly defined magnetoelectric local density of optical states (LDOS) that emerges from cross-coupling of its electric and magnetic resonances~\cite{Kwadrin}.

Although the term LDOS is widely used in nano-optics ~\cite{Snoeks,Chicanne,NovotnyBook}, its precise definition can be elusive~\cite{Joulain,Narayanaswamy}.  Within a bulk (homogeneous) medium, the volume density of electromagnetic modes is a well-defined, spatially-uniform quantity that plays an important role, for example, in the analysis of blackbody emission~\cite{Loudon}. Within structured optical environments though, the density and accessibility of modes can vary with position, hence the term LDOS. The effect of this variation can be seen in the modified lifetimes of electronic emitters near surfaces. Since most common emitters are dominated by electric dipole (ED) transitions, the term LDOS has become almost synonymous with the electric LDOS~\cite{NovotnyBook,Joulain,Narayanaswamy}. 

Recent research has also helped highlight the diversity of electronic transitions in quantum emitters. For example, there has been renewed interest in the magnetic dipole (MD) transitions of lanthanide ions, whose emission rates scale with the magnetic LDOS~\cite{Noginova1, Karaveli2, Schmidt, Rolly, Taminiau2, Dodson, KaraveliNanoLett, Hein}. Nevertheless, these MD transitions are often accompanied by dominant ED transitions that originate from the same excited state. Consequently, the lifetimes of lanthanide emitters are still primarily determined by the electric LDOS~\cite{Drexhage, Chance, Snoeks}. This is also true for transition-metal ions that exhibit spin-forbidden (intersystem) MD zero phonon lines, such as the $^2$E${\rightarrow}^4$A$_2$ transition in Cr$^{3+}$:MgO~\cite{KaraveliACSNano}. These sharp MD lines are accompanied by ED sidebands (mediated by odd-parity phonons) that dominate the emission process at room-temperature.

Interestingly, transition-metal ions can also exhibit spin-allowed MD transitions. These intrasystem transitions are very sensitive to lattice vibrations, which can broaden zero-phonon emission into a band that at room-temperature becomes indistinguishable from its phonon sideband~\cite{Sugano}. As a prototypical example, the broad $^3$T$_2{\rightarrow}^3$A$_2$ emission from divalent nickel (Ni$^{2+}$) ions can range from 1.15 to 1.8 $\mu$m depending on host material and temperature~\cite{Henderson}. Given their broad near-infrared emission, Ni$^{2+}$-doped MgF$_2$ and MgO crystals have been studied for tunable laser applications~\cite{Johnson, Moulton, Iverson2}, and the ED and MD character of their fluorescence has been investigated to varying degrees~\cite{Ferguson,Ralph,Manson,Bird,Wong, Campochiaro}. Cryogenic measurements suggest that phonon sideband emission is predominantly MD (from even-parity phonons) with some potential ED mixing (from odd-parity phonons)~\cite{Henderson,Bird}, whereas the zero-phonon emission is purely MD~\cite{Riley,Mironova-Ulmane}. However, the exact ED and MD contributions to the broad $^3$T$_2{\rightarrow}^3$A$_2$  emission band at room-temperature have not been fully characterized. 

In this Letter, we quantify the ED and MD contributions to the room-temperature, near-infrared emission of Ni$^{2+}$:MgO. Using energy-momentum spectroscopy~\cite{Taminiau2}, we first characterize the intrinsic ED and MD emission rates for the $^3$T$_2{\rightarrow}^3$A$_2$ band, and experimentally demonstrate that MD transitions account for $\sim$50\% of the total intrinsic emission. We then demonstrate that the lifetime of this mixed ED-MD emitter near metal and dielectric interfaces does not solely depend on the electric or magnetic LDOS, but rather follows the combined electromagnetic LDOS. These results show that Ni$^{2+}$:MgO is a quantum emitter that probes the total LDOS much like a thermal emitter, and we discuss the implications for measurements of the local electromagnetic mode density.

\begin{figure}[b] 
  \centering
  \includegraphics[keepaspectratio]{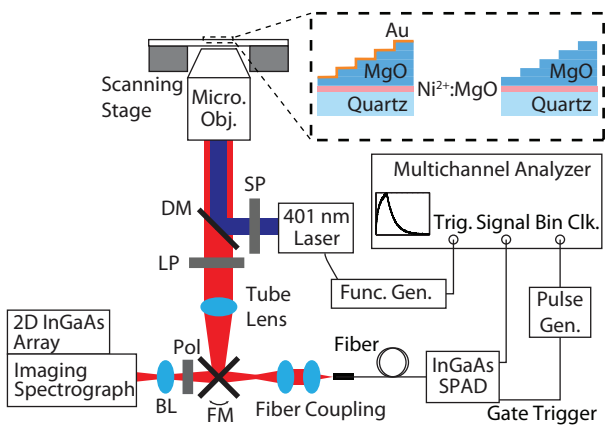}
  \caption{(color online). Schematic of experimental setup. BL: Bertrand lens, DM: Dichroic mirror, LP: Long pass filter, Pol: Linear polarizer, SP: Short pass filter. Flip mirror (FM) switches between the energy-momentum spectroscopy setup (left) and the lifetime measurement setup (right). Inset depicts the MgO steps separating the Ni$^{2+}$:MgO emitter layer from either gold mirror or air interface for the lifetime studies.}
  \label{fig:Schematic}
\end{figure}

The Ni$^{2+}$:MgO emitter layer was fabricated on top of a quartz coverslip by sequential electron beam evaporation of $\sim$18 nm of undoped MgO, $\sim$1 nm of Ni$^{2+}$-doped (1 at.\%) MgO , and $\sim$18 nm undoped MgO. The resulting 37 nm layer thickness was confirmed by ellipsometry and cross-sectional SEM. The sample was subsequently annealed at 1000$^\circ$C for 1h to diffuse the Ni$^{2+}$ ions throughout the layer and thus reduce the doping concentration. For energy-momentum characterization, a 401 nm diode laser was focused to the back-focal-plane of a 1.3 numerical aperture (NA) microscope objective beyond the critical angle in so-called total internal reflection fluorescence mode. As shown in Fig.~\ref{fig:Schematic}, the excitation laser and Ni$^{2+}$ fluorescence were separated by a 775 nm dichroic mirror and a 1064 nm long-pass filter. A rotatable linear polarizer and 100 mm Bertrand lens were used to project polarized momentum-space emission patterns onto the entrance slit of an imaging spectrograph equipped with a 2D InGaAs detector array (Princeton Instruments, IsoPlane SCT 320 with NIRvana). The obtained data has been calibrated for the spectral dependence of the filter, grating and detector system using an intensity-calibrated quartz tungsten halogen lamp (Newport Oriel 63355). 

\begin{figure}[hb] 
  \includegraphics[keepaspectratio]{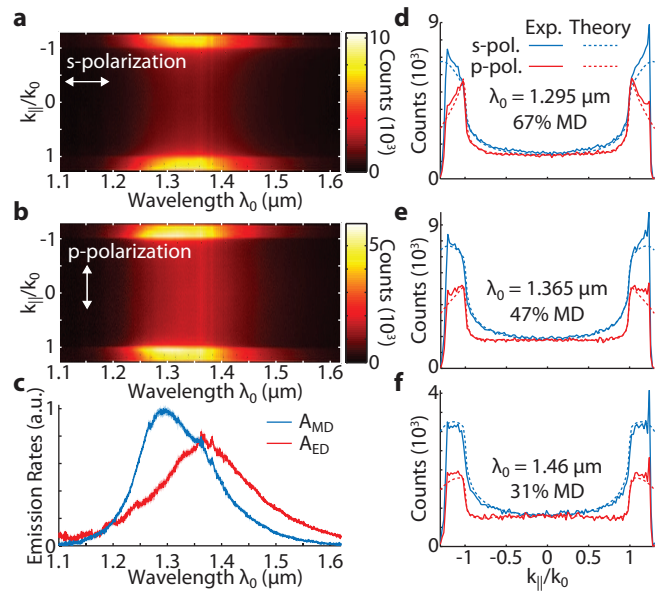}
  \caption{(color online). Quantifying the broad $^3$T$_2{\rightarrow}^3$A$_2$  emission band in Ni$^{2+}$:MgO by energy-momentum spectroscopy. (a,b) Experimental energy-momentum spectra acquired for s- and p-polarization. (c) Spectrally-resolved emission rates, A$_{ED}$ (red line) and A$_{MD}$ (blue line), deduced from fitting analysis together with 95\% confidence intervals (shaded regions). (d-f) Momentum cross sections comparing experimental data (solid lines) with theoretical fits (dashed lines).}
  \label{fig:EnergyMomentum}
\end{figure}
The experimentally acquired energy-momentum spectra for s- and p-polarization are shown in Figs.~\ref{fig:EnergyMomentum}(a) and \ref{fig:EnergyMomentum}(b). We fit the p-polarized momentum cross sections at each wavelength to a linear superposition of the theoretical radiation patterns calculated for isotropic ED and MD emitters located at the center of the 37 nm MgO thin film (n=1.66). As discussed in detail in Ref.~\cite{Taminiau2}, this analysis allows us to obtain the intrinsic spectrally-resolved ED and MD emission rates (Fig.~\ref{fig:EnergyMomentum}(c)) that would be observed in a bulk homogeneous medium. To demonstrate the quality of these fits, Figs.~\ref{fig:EnergyMomentum}(d-f) compare the experimental momentum cross sections with theoretical fits at three representative wavelengths: $\lambda_0$ = 1.295 $\mu$m where the MD emission rate peaks and accounts for 67\% of the total emission rate; $\lambda_0$ = 1.365 $\mu$m where the ED emission rate peaks but MD emission still accounts for 47\%; and $\lambda_0$ = 1.46 $\mu$m where ED emission accounts for 69\% of the total emission. Note that the $\sim$400 cm$^{-1}$ energy separation between the MD and ED emission rate maxima is consistent with the average phonon energy in MgO~\cite{Peckham}, suggesting that ED and MD transitions are distinguished by a single phonon process. 

Most importantly for this study, the rates shown in Fig.~\ref{fig:EnergyMomentum}(c) indicate that this broadband emission has a strongly mixed ED and MD character. Integrating over these spectrally-resolved rates, we find that MD transitions account for approximately half (50.4 $\pm$ 2.5\%) of the total emission rate for the $^3$T$_2{\rightarrow}^3$A$_2$ band. To demonstrate that the $^3$T$_2{\rightarrow}^3$A$_2$ transition depends on the total electromagnetic LDOS and, thus, that Ni$^{2+}$:MgO serves as a quantum mechanical probe of the local electromagnetic mode density, we have performed lifetime experiments to study the modified spontaneous emission rates near planar interfaces. 

For the lifetime study, we varied the distance of the Ni$^{2+}$:MgO thin film to both a gold mirror and an air interface, and acquired time decay traces of its photoluminescence. For this purpose, different thickness spacers were fabricated by consecutive evaporation of undoped MgO while masking parts of the sample to achieve $\sim$2 mm wide steps with heights ranging from $\sim$25 nm to $\sim$475 nm. Then, a gold mirror was deposited on a portion of each step by evaporating a 5 nm Ti adhesion layer followed by 200 nm of Au. Schematic illustrations of the final structures are shown in the inset of Fig.~\ref{fig:Schematic}. The sample was excited, and emission collected, with a 20x (0.75 NA) microscope objective under confocal illumination. The 401 nm diode laser was modulated using a function generator to obtain 5.26 ms pulses at a 38 Hz repetition rate. As shown in Fig.~\ref{fig:Schematic}, the collected emission was focused into a fiber coupled InGaAs/InP near-infrared single photon avalanche photodiode (SPAD) detector module (Micro Photon Devices)~\cite{Tosi}.  Histograms of photon arrival times were obtained using a multichannel analyzer (Stanford Research Systems, SR430) to count the photon events at 12.5 $\mu$s time-bin intervals. (See Supplemental Material for more details ~\cite{SuppMat}.)
 \begin{figure}[b] 
  \centering
  \includegraphics[keepaspectratio]{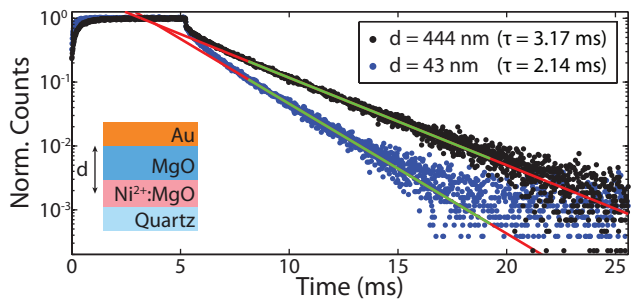}
  \caption{(color online). Examples of time-resolved photoluminescence data and fits used to determine the excited state lifetime of Ni$^{2+}$:MgO. Selected data points highlight different lifetimes observed close to (blue) and far from (black) the gold mirror.  Green line segments show the temporal regions used for single-exponential fits to the long-lived $^3$T$_2$ state.}
  \label{fig:TimeTrace}
\end{figure}
 \begin{figure}[b] 
  \centering
  \includegraphics[keepaspectratio]{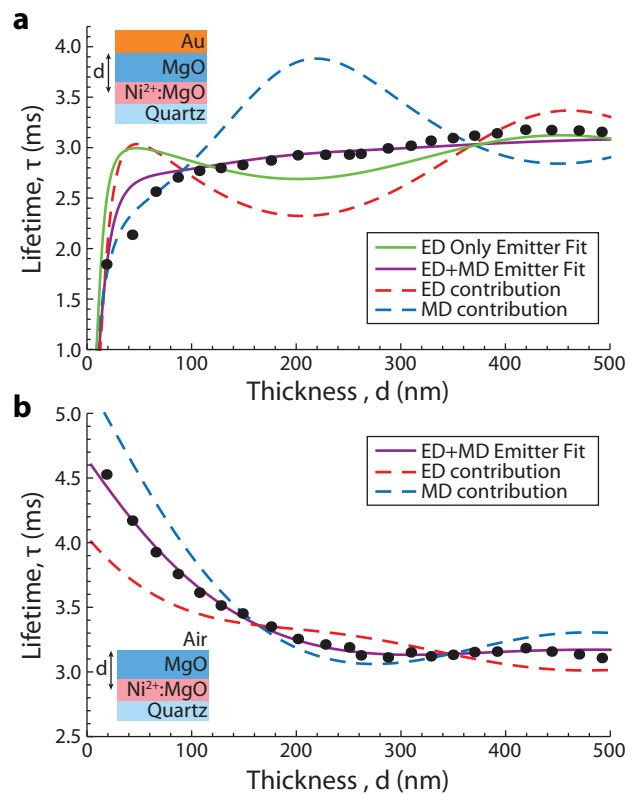}
  \caption{(color online). Lifetime data acquired for $^3$T$_2$ state of Ni$^{2+}$:MgO near (a) a gold mirror and (b) an air interface together with fits to the purely electric LDOS (green line) and mixed electromagnetic LDOS (purple line). For (a) the gold mirror case, a least squares fit shows that the data follows the electromagnetic LDOS obtained for a 52.4 $\pm$ 8.2\% MD emitter with a 75.6\% quantum yield. For (b) the air interface case, a least squares fit shows that the data follows the electromagnetic LDOS obtained for a 56.9 $\pm$ 4.8\% MD emitter with a 77.8\% quantum yield. For completeness, we also plot the electric (red dashed) and magnetic (blue dashed) contributions to the electromagnetic LDOS, i.e. the variations expected for ED and MD transitions which are then averaged according to the fit percentages to obtain the purple curve.}
  \label{fig:Lifetimes}
\end{figure}

Example time decay traces for two different spacer layer thicknesses are shown in Fig.~\ref{fig:TimeTrace} together with fit results to a single exponential decay. Although an effort has been made to minimize the Ni$^{2+}$ ion concentration in the emitter layer, the initial fast decay observed immediately after the excitation pulse may be due to concentration based effects, which result in distinct Ni$^{2+}$ ion populations with different decay rates~\cite{Tonucci}, as well as short-lived excited state luminescence~\cite{Payne}. Therefore, as highlighted by green lines in Fig.~\ref{fig:TimeTrace},  all fits were performed over the same time region between 3 and 14 ms after the excitation pulse ends to isolate the excited state lifetime of the long-lived $^3$T$_2$ state.

Figure~\ref{fig:Lifetimes} shows the measured lifetimes as a function of distance \textit{d} from the center of the Ni$^{2+}$:MgO thin film to the gold and air interfaces. Note that the lifetimes do not exhibit large oscillations, as would be expected for dominantly ED or MD transitions. Instead, Fig.~\ref{fig:Lifetimes}(a) shows a steep rise in the lifetime followed by a gentle increase as the emitter-mirror separation is increased, whereas for increasing distances from the air interface, Fig.~\ref{fig:Lifetimes}(b) shows a decrease toward a plateau. In the literature~\cite{Chance,Snoeks}, lifetime variations are commonly used to infer the quantum efficiency by fitting the distance-dependent lifetime $\tau(d)$ to the following equation
\begin{equation}
\tau(d)=\tau_0 (1-q(1-\tilde{\Gamma}(d)))^{-1}
\label{eq1}
\end{equation}
where the fit parameters \textit{q} and $\tau_0$ correspond to the emitter's quantum efficiency and lifetime in a homogenous medium, respectively, and where $\tilde{\Gamma}(d)\equiv{\Gamma(d)}/{\Gamma_0}$ is the emission rate normalized to the homogeneous case. For isotropic ED emitters~\cite{Chance,Drexhage,Kunz,Snoeks}, this rate is equivalent to the normalized electric LDOS, $\tilde{\rho}^{E}(d)$~\cite{Joulain,NovotnyBook}
\begin{equation}
\tilde{\Gamma}^{Iso}_{ED}(d)\equiv\tilde{\rho}^{E}(d)=\frac{Im\{Tr[G^E({\bf r},{\bf r})]\}}{Im\{Tr[G^E_0({\bf r},{\bf r})]\}}
\label{eq2}
\end{equation}
where $G^E(r,r)$ and $G^E_0(r,r)$ are the electric field Green's Dyadic functions for the inhomogeneous system and homogeneous reference.  Likewise, a similar expression can be derived for isotropic MD emitters in terms of the magnetic LDOS, $\tilde{\rho}^{M}(d)$, and the Green's Dyadic for the magnetic field $G^H(r,r)$~\cite{Chance,Joulain,Karaveli2,Taminiau2}
\begin{equation}
\tilde{\Gamma}^{Iso}_{MD}(d)\equiv\tilde{\rho}^{M}(d)=\frac{Im\{Tr[G^H({\bf r},{\bf r})]\}}{Im\{Tr[G^H_0({\bf r},{\bf r})]\}}.
\label{eq3}
\end{equation}
The explicit expressions used for $\tilde{\rho}^{E}(d)$ and $\tilde{\rho}^{M}(d)$, including how they are averaged over the emitter's spectral range, are provided in the Supplemental Material~\cite{SuppMat}.

The lifetime values shown in Fig.~\ref{fig:Lifetimes} cannot be fit to purely ED or purely MD emitters. Although one might assume that the relative absence of oscillations could be attributed to a low quantum efficiency, this cannot account for the large variations observed near the gold and air interfaces. To demonstrate this point, the green line in Fig.~\ref{fig:Lifetimes}(a) shows the best least squares fit obtained using Eqs.~(\ref{eq1})-(\ref{eq2}) and assuming an isotropic pure ED emitter with $\tau_0$ and \textit{q} as fit parameters. This ED fit clearly does not follow the observed data, especially at short distances from the gold mirror, for which ED emission is inhibited but our measurements show a marked decrease in lifetime. (See Supplemental Figure S2 for similar fits to a purely MD emitter ~\cite{SuppMat}.)

The measured lifetimes can be fit to a superposition of isotropic ED and MD emission. To do so we set $\tilde{\Gamma}(d)\equiv a_{MD}\tilde{\Gamma}^{Iso}_{MD}(d)+(1-a_{MD})\tilde{\Gamma}^{Iso}_{ED}(d)$, where $a_{MD}$ is the MD percentage of total emission that we use as a fit variable parameter together with $\tau_0$. To reduce the number of fit parameters, we leverage the fact that the quantum efficiency can be approximated by: $q=\tau_0/\tau_{rad}$, where  $\tau_{rad}$=3.6 ms is the intrinsic radiative lifetime of the $^3$T$_2{\rightarrow}^3$A$_2$ transition of Ni$^{2+}$:MgO inferred from temperature dependent measurements~\cite{Iverson2}. The resulting fits are shown by the purple lines in Figs.~\ref{fig:Lifetimes}(a) and~\ref{fig:Lifetimes}(b). For the gold case, we obtain $\tau_0$ = 2.72 $\pm$ 0.06 ms and $a_{MD}$ = 52.4 $\pm$ 8.2\%, and these results agree well with the experimental lifetime data. For completeness, the dashed lines in Fig.~\ref{fig:Lifetimes} show the calculated lifetime variations associated with the ED and MD contributions. Similar fitting analysis for the air case yields $\tau_0$ = 2.81 $\pm$ 0.01 ms and $a_{MD}$ = 56.9 $\pm$ 4.8\%. Within the 95\% confidence intervals, the $a_{MD}$ values obtained from both sets of lifetime measurements agree well with the earlier energy-momentum characterization. Indeed, without any fitting parameters, the intrinsic rates shown in Fig. 2(c) can be used to directly predict the observed lifetime variations (see Supplemental Figure S3 \cite{SuppMat}). Thus, our experiments confirm that light emission from the $^3$T$_2{\rightarrow}^3$A$_2$ band is due to nearly equal ED and MD contributions (i.e. $a_{MD}\approx$ 50\%). 

For the special case where the intrinsic ED and MD rates are identical (i.e. $a_{MD}=$ 50\%), the normalized rate $\tilde{\Gamma}$ scales with the combined electromagnetic LDOS, 
\begin{equation}
\tilde{\rho}^{EM}(d)=\frac{Im\{Tr[G^E({\bf r},{\bf r})]+Tr[G^H({\bf r},{\bf r})]\}}{Im\{Tr[G^E_0({\bf r},{\bf r})]+Tr[G^H_0({\bf r},{\bf r})]\}}
\label{eq4}
\end{equation}
which can be derived from Eqs.~(\ref{eq2})-(\ref{eq3}) by noting that $Im\{Tr[G^E_0({\bf r},{\bf r})]\}=Im\{Tr[G^H_0({\bf r},{\bf r})]\}$ in a bulk medium ~\cite{Joulain,Narayanaswamy}. Therefore, having identified an emitter with roughly equal ED and MD emission rates, its lifetime can be used as a probe of  $\tilde{\rho}^{EM}$. Unlike $\tilde{\rho}^{E}$ and $\tilde{\rho}^{M}$, this quantity is less sensitive to far-field interference, i.e. the spatial variations between electric and magnetic field maxima.  However, it is still very sensitive to near-field phenomena. Indeed, in Fig.~\ref{fig:Lifetimes}, we see the $\sim$3 ms lifetime observed far from both interfaces increase by $\sim$50\% near the air surface and decrease by $\sim$40\% near the gold mirror. These large changes result from the decreased mode density near the low index air and the increased contributions of surface modes near the gold film. 

In conclusion, we have identified a room-temperature, solid-state emitter Ni$^{2+}$:MgO that probes the combined electric and magnetic LDOS, i.e. the electromagnetic LDOS experienced by an incoherent sum of ED and MD transitions. Interestingly, this electromagnetic LDOS is the same quantity probed by thermal emission~\cite{Joulain}. In this context, these results complement recent work on thermal radiation scanning tunnelling microscopy (TRSTM)  \cite{Wilde,Jones,Babuty} and may provide a new experimental system with which to  study near-field radiative heat transfer~\cite{JoulainSurfSci}. Similar to recent scanning lifetime measurements where ED emitters were used to map the electric LDOS \cite{Frimmer}, one can envision depositing Ni$^{2+}$:MgO on a near-field fiber probe and using the $^3$T$_2$ excited state lifetime as a direct measure of the electromagnetic LDOS. Since the strongly mixed ED-MD transitions in this system mimic thermal emission, such measurements could allow one to isolate and investigate thermal-like radiation effects without the need to heat samples nor minimize other thermal transport channels (i.e. conduction and convection). Lifetime measurements with these isotropic emitters also probe all components of the electromagnetic LDOS, unlike TRSTM scattering measurements which probe a projection of the LDOS along the tip axis~\cite{Wilde}. Additionally, by examining lifetimes rather than intensities and working at near-infrared rather than mid-infrared wavelengths, such measurements could substantially reduce experimental complexity.

More generally, the study of these and other multipolar emitters may help broaden the range of electromagnetic phenomena that can be accessed with electronic systems. Having identified a thermal-like emitter with near-equal incoherent ED and MD contributions, an open question remains as to whether this or other solid-state materials can be engineered to exhibit a coherent superposition of near-equal ED and MD transitions. Such emitters could help realize atomic analogues to the recent phenomena explored with macroscopic ED and MD scatterers~\cite{Geffrin, FuNatComm, Person,Etxarri2,Sersic1}.

The authors thank S. Cueff, C. M. Dodson, C. Hargus, M. Jiang, J. A. Kurvits, D. Paine, A. Peterson, and T. H. Taminiau for helpful discussions. Financial support for this work was provided by the Air Force Office of Scientific Research (MURI award FA-9550-12-1-0488) and the National Science Foundation (CAREER award EECS-0846466).

\bibliographystyle{aipnum4-1}

\end{document}